\documentclass{article}

\PassOptionsToPackage{numbers, compress}{natbib}
\PassOptionsToPackage{dvipsnames,svgnames,x11names,HTML}{xcolor}
\usepackage{tech2025}

\usepackage[utf8]{inputenc} 
\usepackage[T1]{fontenc}    
\usepackage[colorlinks=true,citecolor=SkyBlue]{hyperref}       
\usepackage{url}            
\usepackage{booktabs}       
\usepackage{amsfonts}       
\usepackage{nicefrac}       
\usepackage[nopatch=footnote]{microtype}
\usepackage{graphicx}

\usepackage{setspace}
\usepackage[most]{tcolorbox}
\usepackage[colorinlistoftodos]{todonotes}
\usepackage{array}
\usepackage{booktabs}
\usepackage{tabularx}
\usepackage{placeins}
\usepackage{silence}
\usepackage{xcolor}
\usepackage{colortbl}
\usepackage{multirow,makecell}
\usepackage[capitalise,noabbrev]{cleveref}
\usepackage{epigraph}
\usepackage{pifont}
\usepackage{ragged2e}
\usepackage{arydshln}
\usepackage{tikz}
\usetikzlibrary{decorations.shapes}
\usepackage{amsfonts}
\usepackage{amssymb}
\usepackage{enumitem}
\usepackage[normalem]{ulem}
\usepackage{hypcap}
\usepackage{capt-of}
\usepackage{fontawesome}
\usepackage{bm}
\usepackage{atbegshi}
\usepackage{nameref}
\usepackage{colortbl}
\usepackage{subfigure}
\usepackage{xspace,mfirstuc,tabulary}
\usepackage{newunicodechar}
\usepackage{atbegshi}
\usepackage{mdframed}
\usepackage{algorithm}
\usepackage[noend]{algpseudocode}
\usepackage{soul}
\usepackage{bbm}
\usepackage{rotating}
\usepackage[outline]{contour}
\usepackage{ifthen}
\usepackage{framed}
\usepackage{amsmath}
\usepackage{mathtools}

\usepackage{subfigure}
\usepackage{wrapfig}
\usepackage{caption}
\usepackage{float}
\usepackage{tcolorbox}
\usepackage{xspace}
\usepackage{pifont}
\usepackage[misc,geometry,weather]{ifsym}

\DeclareCaptionLabelSeparator{pipe}{\ $\mid$ }

\newcommand{\GitHubLogo}{\raisebox{-0.15em}{\includegraphics[height=1.1em]{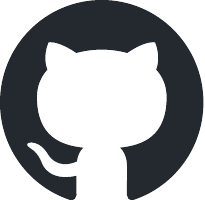}}}

\definecolor{OliveGreen}{rgb}{0.0, 0.5, 0.0}
\definecolor{NavyBlue}{rgb}{0.0, 0.0, 0.5}
\definecolor{prompttitlebg}{HTML}{E6EEF9} 
\definecolor{promptframe}{HTML}{B0C4DE}   
\definecolor{promptbg}{HTML}{F8F9FA}      

\newtcolorbox{prompt}[1]{
    breakable,
    colback=promptbg,
    colframe=promptframe,
    boxrule=1pt,
    arc=2mm,
    title={\textsc{#1}}, 
    fontupper=\ttfamily\small,
    fontlower=\ttfamily\small,
    colbacktitle=prompttitlebg, 
    coltitle=black, 
    parbox=false,  
    text width=\linewidth-8mm,  
    top=2mm,    
    bottom=4mm,
    left=4mm,
    right=4mm,
    boxsep=1mm,
    toptitle=1.5mm, 
    bottomtitle=1.5mm, 
    enhanced,
    shadow={.5mm}{-.5mm}{0mm}{black!15},
    before upper={\setlength{\parindent}{0pt}\setlength{\parskip}{6pt}}, 
    middle=2mm, 
    after skip=\baselineskip, 
    before skip=\baselineskip 
}

\makeatletter
\def\icmldate#1{\gdef\@icmldate{#1}}
\icmldate{\today}
\makeatother
\definecolor{bigaired}{RGB}{156, 0, 0}
\definecolor{uclablue}{RGB}{39, 116, 174}

\definecolor{darkred}{RGB}{200, 0, 0}
\definecolor{darkblue}{RGB}{0, 0, 200}
\definecolor{blue}{RGB}{0, 0, 250}

\definecolor{light}{RGB}{225, 250, 250}
\definecolor{lightgray}{RGB}{0.9, 0.9, 0.9}
\definecolor{lightred}{RGB}{250, 200, 200}
\definecolor{lightblue}{RGB}{210, 220, 250}

\definecolor{doderblue}{RGB}{30, 144, 255}
\definecolor{select}{RGB}{222, 235, 247}
\definecolor{unselect}{RGB}{247, 207, 206}

\definecolor{lightgrey}{RGB}{247, 247, 247}
\definecolor{myblue}{RGB}{39,116,174}

\hypersetup{citecolor=uclablue, linkcolor=uclablue, urlcolor=bigaired}

\newenvironment{bigaiabstract}{
  \begin{tcolorbox}[
    colback=lightgrey,
    colframe=white,
    boxrule=0pt,
    arc=10pt,
    left=16pt,
    right=16pt,
    top=12pt,
    bottom=12pt,
    width=\textwidth,
    enlarge left by=0mm,
    before skip=10pt,
    after skip=10pt
  ]
  \normalsize
}{
  \end{tcolorbox}
}

\makeatletter
\newcommand*{\hyperlinkcite}[1]{\hyper@link{cite}{cite.#1}}
\makeatother

\usepackage{amsmath,amsfonts,bm}

\def\eqref#1{equation~\ref{#1}}

\def\1{\bm{1}}

\DeclareMathAlphabet{\mathsfit}{\encodingdefault}{\sfdefault}{m}{sl}
\SetMathAlphabet{\mathsfit}{bold}{\encodingdefault}{\sfdefault}{bx}{n}

\makeatletter
\DeclareRobustCommand\onedot{\futurelet\@let@token\@onedot}
\def\@onedot{\ifx\@let@token.\else.\null\fi\xspace}

\makeatother

\runningtitle{Evaluating Auditory Motion Perception in Audio LLMs}
    
\title{AudioMotionBench: Evaluating Auditory Motion Perception in Audio LLMs}

\author{
Zhe Sun$^{1}$ \quad
Yujun Cai$^{2}$\thanks{Corresponding author.} \quad
Jiayu Yao$^{3}$ \quad
Yiwei Wang$^{4}$
\\[1mm]
\normalfont
$^{1}$Wuhan University \quad
$^{2}$University of Queensland \\
$^{3}$University of Chinese Academy of Sciences \quad
$^{4}$University of California, Merced
\\[0.5mm]
\texttt{bubbachuck33333@gmail.com} \quad
\texttt{yujun.cai@uq.edu.au}
}

\begin{document}

\maketitle

\begin{center}
\setlength{\tabcolsep}{7pt}
\begin{tabular}{cll}
    \GitHubLogo & \textbf{Code:} & \url{https://github.com/sunzhe12/AudioMotionBench}\\
\end{tabular}
\end{center}

\begin{bigaiabstract}
\begin{center}
    \LARGE\bf Abstract
\end{center}

Large Audio-Language Models (LALMs) have recently shown impressive progress in speech recognition, audio captioning, and auditory question answering. Yet, whether these models can perceive spatial dynamics, particularly the motion of sound sources, remains unclear.  In this work, we uncover a systematic motion perception deficit in current ALLMs. To investigate this issue, we introduce \textbf{AudioMotionBench}, the first benchmark explicitly designed to evaluate auditory motion understanding. AudioMotionBench introduces a controlled question-answering benchmark designed to evaluate whether Audio-Language Models (LALMs) can infer the direction and trajectory of moving sound sources from binaural audio. Comprehensive quantitative and qualitative analyses reveal that current models struggle to reliably recognize motion cues or distinguish directional patterns. The average accuracy remains below 50\%, underscoring a fundamental limitation in auditory spatial reasoning. Our study highlights a fundamental gap between human and model auditory spatial reasoning, providing both a diagnostic tool and new insight for enhancing spatial cognition in future Audio-Language Models.

\end{bigaiabstract}

\section{Introduction}
\label{sec:introduction}

\begin{figure}[t]
    \centering
    \includegraphics[width=\linewidth]{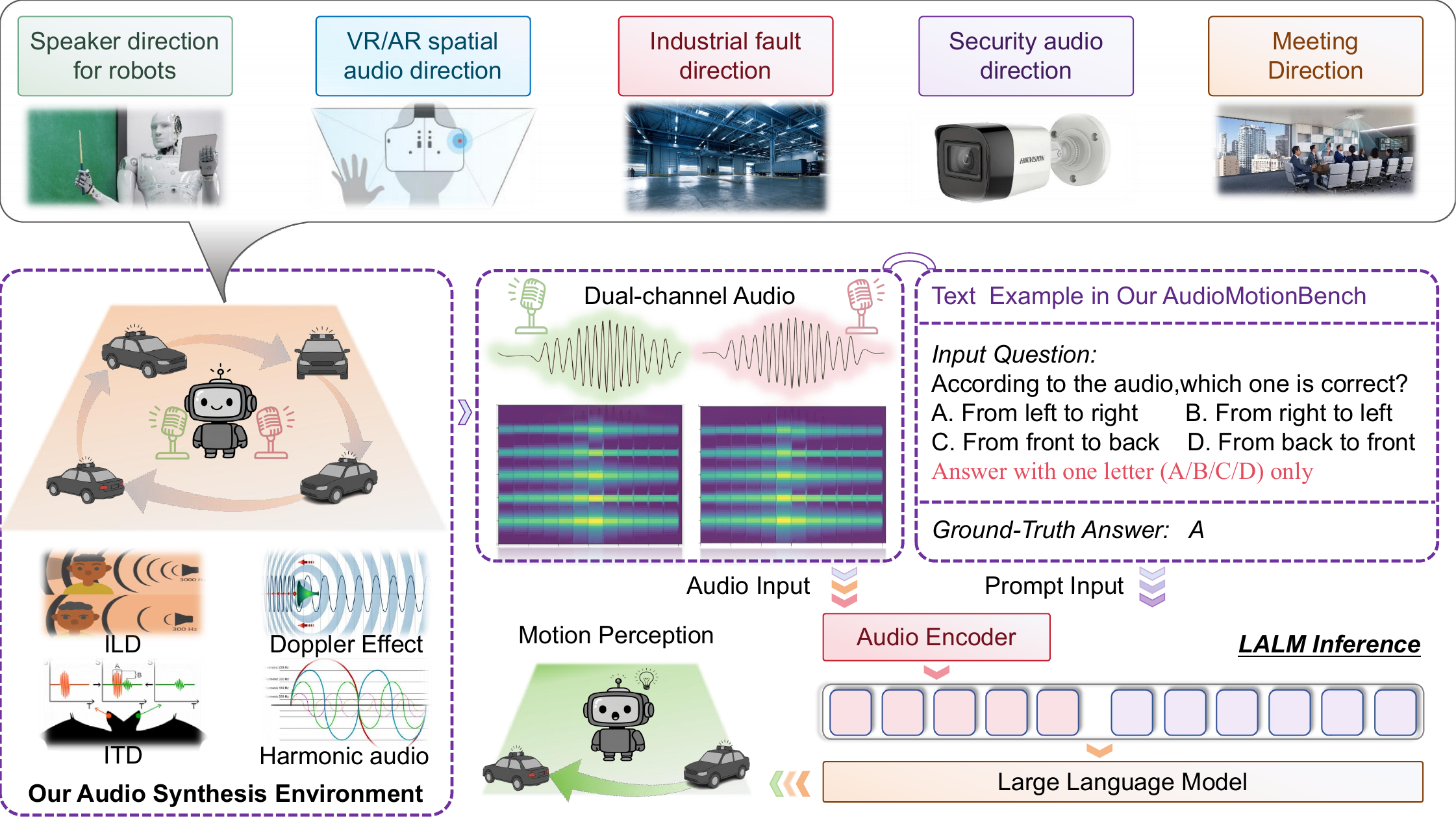}
    \caption{\textbf{Overview of AudioMotionBench and the audio motion perception evaluation pipeline},The benchmark simulates controlled binaural audio with explicit motion cues to model moving sound sources in realistic environments. Dual-channel audio inputs are encoded and paired with structured multiple-choice questions that require reasoning about sound source direction and trajectory. }
    \label{fig:ampbench_overview}
\end{figure}
Sound is more than a sequence of acoustic events—it is a spatial and dynamic signal that encodes how the world moves around us. From the subtle Doppler shift of a passing car to the changing loudness of a speaker walking across a room, motion in sound provides a continuous stream of physical cues that anchor our auditory experience in space. Humans effortlessly interpret these cues to understand not only what is happening, but also where it occurs and how it evolves over time.

Recent Large Audio-Language Models (LALMs), such as Qwen2-Audio-7B, Aero-1-Audio, Voxtral-Mini-3B-2507,and Kimi-Audio-7B \cite{chu2024qwen2,liu2025voxtral,mandel2023aero,ding2025kimi}, have achieved remarkable progress in understanding and reasoning over speech, music, and environmental sounds. Yet a fundamental aspect of auditory intelligence remains largely unexplored: the perception of motion. While current models perform well on recognizing sound types and semantic content, existing audio-language tasks rarely require them to reason about the dynamic movement and trajectory of sound sources in space.

Human auditory intelligence, is profoundly spatial. The brain continuously fuses interaural time differences (ITD), interaural level differences (ILD), Doppler shifts, and spectral filtering to reconstruct the three-dimensional motion of sound sources \cite{chu2025learning}. Such spatial integration enables humans to localize, track, and anticipate motion even in noisy or reverberant conditions. In contrast, most existing LALMs are trained on single-channel or weakly spatialized datasets that emphasize semantic comprehension but neglect spatial reasoning, leading to a fundamental question: Can modern Audio-Language Models truly perceive motion?

To answer this question, we introduce the \textbf{Audio Motion Perception Benchmark (AudioMotionBench)} (Figure~\ref{fig:ampbench_overview}), the first benchmark specifically designed to evaluate motion understanding in LALMs. AudioMotionBench provides physically simulated binaural audio clips under controlled spatial conditions, paired with structured question-answering tasks that probe directionality and trajectory reasoning across multiple noise levels. This benchmark moves beyond static sound recognition to test whether models can infer spatial dynamics directly from the waveform. Through extensive zero-shot evaluation of five state-of-the-art systems, we reveal a consistent and systematic motion perception deficit: models fail to correctly recognize motion cues and directional patterns, achieving below 50\% accuracy on average. This benchmark thus exposes a key limitation in current audio-language understanding and establishes a diagnostic foundation for developing spatially aware auditory models.

We obtained the following research results through controlled experiments and systematic benchmark evaluations:
\begin{itemize}[noitemsep, topsep=0pt]
    \item We uncover and formalize the overlooked problem of motion perception in Large Audio-Language Models, showing that existing systems can recognize semantic content but fail to perceive how sounds move through space, which is a core component of auditory intelligence long ignored in LALMs. 
    \item We propose \textbf{AudioMotionBench}, a controlled and reproducible benchmark for evaluating spatial reasoning from binaural audio. It provides binaural simulations under well-defined spatial parameters and structured question–answering tasks that directly test models’ understanding of auditory motion.
    \item Through extensive zero-shot experiments across representative LALMs, we reveal systematic motion perception deficits in spatial reasoning. These findings establish a diagnostic foundation for developing LALMs with genuine spatial awareness.
\end{itemize}

\section{Related Work}
\label{sec:related}

Recent work on large audio-language models (LALMs) has made impressive strides in semantic tasks such as transcription, summarization, and audio-based question answering. However, most efforts focus on what is being said rather than where it happens or how sound sources move. This semantic bias leaves a blind spot in spatial and dynamic audio understanding, motivating the need for dedicated benchmarks and methods targeting motion and spatial reasoning in audio.  

\subsection{Large Audio-Language Models}

Large Audio-Language Models (LALMs) extend large language models by incorporating audio perception and reasoning capabilities. These models typically employ pretrained audio encoders (e.g., WavLM\cite{chen2022wavlm}) to extract high-level acoustic representations, which are then fused with a language-model backbone and fine-tuned for downstream audio-text tasks \cite{su2025audio}. Recent surveys show the field of LALMs is growing rapidly, covering model architecture, multimodal fusion strategies, and evaluation benchmarks \cite{yang2025towards}.

Several recent systems exemplify this trend\cite{ding2025kimi,xie2025audio,hurst2024gpt,ghosh2024gama,goel2025audio}. For example, the Qwen-Audio model unified pre-training across a large set of audio tasks and domains and achieved strong zero-shot performance on audio understanding benchmarks\cite{chu2023qwen}. More recently, the Qwen2.5-Omni model pushes toward full multimodal input including audio, video and image, thus marking a transition from pure audio-language to richer modality integration \cite{xu2025qwen2}. Meanwhile, Xiaomi’s MiMo-Audio model—trained on over 100 million hours of audio and using a unified audio-text next-token objective—demonstrates the push for scale and generality in audio-language modelling \cite{coreteam2025mimoaudio}.

Although these advances have considerably broadened the landscape of auditory reasoning, existing LALMs still focus primarily on semantic comprehension. Most current models describe what happens in a sound but fail to capture where it occurs or how it evolves through space and time. However, none of these works explicitly probe \textbf{directional perception} or \textbf{motion continuity}—two aspects that are central to spatial auditory intelligence. This fundamental gap motivates our work on spatial and motion reasoning in LALMs.

\subsection{Spatial and Multimodal Perception}

Spatial and multimodal perception in the audio domain addresses the ``where'' and ``how'' of sound: not only what source is present, but also its direction, distance, motion, and relationship with visual/contextual cues. Recent research has begun to explore spatial audio understanding and generation, highlighting the need for models that go beyond semantic recognition and reason about spatial and multimodal context \cite{zhu2025asaudio,sarabia2023spatial}.

For example, the ELSA model (Embeddings for Language and Spatial Audio) learns spatially-aware audio and text embeddings via contrastive learning over spatial configurations, showing that standard audio‐text models lack spatial awareness \cite{devnani2024learning}. The BAT model (Learning to Reason about Spatial Sounds with Large Language Models) demonstrates how a spatial audio encoder can be combined with a language model to answer spatial sound reasoning questions \cite{zheng2024bat}. More recently, SALM (Spatial Audio Language Model) integrates structured audio embeddings separating semantic and spatial components, enabling zero-shot direction classification and editing of spatial audio scenes \cite{hu2025salm}. These works illustrate a growing trend toward spatially-grounded audio-language modelling.

Meanwhile, Sarabia et al. introduced Spatial LibriSpeech, a dataset augmented with multi-channel spatial audio and annotations for source position, direction and room acoustics \cite{sarabia2023spatial}. Other works examine spatially aligned audio–video generation benchmarks such as SAVGBench which focus on spatialized audio–visual coherence \cite{shimada2024savgbench}. 

Despite this progress, most LALMs still rely on single-channel or weakly spatialized datasets and rarely model motion trajectories or multimodal cues (e.g., visual context) in their training. The field of spatial audio generation further shows that tasks such as moving sound source synthesis, multichannel rendering, and binaural reasoning remain under-explored \cite{zhu2025asaudio, liu2025text2move, groger2025evaluation}. Consequently, spatial and multimodal perception remains a bottleneck for truly embodied audio-language intelligence.

\subsection{Benchmarks and Datasets}

Several benchmarks have been developed to evaluate audio–language models. For instance, the AudioBench benchmark provides a universal evaluation framework covering eight tasks and 26 datasets targeting LALMs \cite{wang2024audiobench}. The MMAU benchmark further advances this by offering 10{,}000 audio clips paired with human-annotated questions and answers spanning speech, environmental sounds and music, and evaluates models on 27 distinct skills including reasoning and information extraction \cite{sakshi2024mmau}. Another recent benchmark, AIR-Bench, is designed to assess LALMs’ interaction capabilities via generative comprehension across speech, natural sounds and music, with both foundation and chat tasks \cite{yang2024air}.  

While these resources provide valuable coverage of semantic and reasoning tasks, they largely overlook the continuous spatial and motion aspects of auditory scenes.As summarized in Table~\ref{tab:benchmark_comparison}, existing audio–language benchmarks primarily emphasize semantic understanding and broad task coverage, while largely overlooking controlled auditory motion perception. In contrast, AudioMotionBench is explicitly designed to isolate continuous spatial cues in binaural audio under physically controlled conditions, minimizing semantic and acoustic confounds. This makes AudioMotionBench complementary to prior benchmarks and uniquely suited for probing motion-specific auditory reasoning.

\begin{table}[t]
\centering
\small
\begin{tabular}{lcccc}
\toprule
\textbf{Benchmark} 
& \textbf{Audio Modality} 
& \textbf{Task Diversity} 
& \textbf{Continuous Spatial Cues} 
& \textbf{Real-World Audio} \\
\midrule
AudioBench \cite{wang2024audiobench} 
& Mono / Multi 
& High 
& \ding{55} 
& \ding{51} \\

MMAU \cite{sakshi2024mmau} 
& Mono 
& High 
& \ding{55} 
& \ding{51} \\

AIR-Bench \cite{yang2024air} 
& Mono 
& Medium 
& \ding{55} 
& \ding{51} \\

\midrule
\textbf{AudioMotionBench} 
& \textbf{Binaural} 
& Medium 
& \ding{51} 
& \ding{55} \\
\bottomrule
\end{tabular}
\caption{Comparison between AudioMotionBench and existing audio--language benchmarks. Existing benchmarks emphasize broad task coverage and real-world audio, while AudioMotionBench focuses on controlled auditory motion perception through continuous spatial cues in binaural signals.}
\label{tab:benchmark_comparison}
\end{table}

\section{The AudioMotionBench Framework}
\label{sec:Benchmark}

Building upon the limitations identified in prior benchmarks, AudioMotionBench aims to provide a systematic evaluation of auditory motion perception. Whereas previous datasets emphasize semantic or multimodal understanding, AudioMotionBench isolates motion as an independent perceptual dimension and measures whether large audio–language models (LALMs) can infer direction and continuity purely from binaural cues. The following subsections describe its task formulation, synthesis pipeline, and dataset composition in detail.

\subsection{Task Formulation}

AudioMotionBench evaluates a model’s ability to perceive and reason about moving sound sources using only auditory input. We formulate this as an auditory motion question–answering task in which a model must infer the direction and trajectory of a moving sound source based on a binaural waveform and a natural-language query. 

Formally, each sample consists of a binaural audio clip $x=(x_L,x_R)$ synthesized under controlled spatial configurations and paired with a motion-related question $q$. The model $f_{\theta}$ predicts an answer $y$ from a constrained label set. Two formats are supported: a multiple-choice task that selects the correct direction among predefined candidates, and a true-or-false verification task that checks whether a textual description matches the actual motion. The task can be formalized as a constrained label prediction problem:
\begin{equation}
\begin{gathered}
y = f_{\theta}(x, q), \\
y \in 
\begin{cases}
\mathcal{Y}_{\mathrm{MC}}=\{\mathrm{A,B,C,D}\}, &\text{multiple choice},\\[2pt]
\mathcal{Y}_{\mathrm{TF}}=\{\mathrm{TRUE,FALSE}\}, &\text{true/false}.
\end{cases}
\end{gathered}
\end{equation}

Here $f_{\theta}$ denotes the audio–language model under evaluation and $y$ represents the predicted label corresponding to auditory motion understanding. Performance is measured by classification accuracy, following standard benchmark practice, providing a transparent and comparable metric of motion perception and reasoning ability.

\subsection{Physically-Grounded Audio Synthesis}

To ensure interpretability and reproducibility, AudioMotionBench employs an analytical binaural rendering model instead of black-box HRTF convolution. Each instance simulates a single sound source moving in a controlled space, with the listener fixed at the center. In detail, the sound field is defined within a square area of $6\,\text{m}\times6\,\text{m}$, where the $y$-axis points forward and the $x$-axis points to the right. The listener is positioned at the center $(3.0, 3.0)$, corresponding to the midpoint of the space. Eight canonical directions are defined along the horizontal plane: front (N), back (S), left (W), right (E), and the four diagonals (NE, NW, SE, SW). Each sound trajectory corresponds to a straight-line motion between two of these positions, such as NE$\!\rightarrow$SE (right-front to right-back) or NW$\!\rightarrow$SW (left-front to left-back). This setup provides a physically interpretable coordinate system that supports reproducible and directionally balanced sound motion generation.

For every frame in a trajectory, binaural signals are computed from the instantaneous distance between the source and each ear. ITD is obtained from propagation delay differences, ILD from inverse-square distance attenuation, and Doppler shift from the radial velocity component. To reproduce front–back distinction, a direction-dependent low-pass filter is applied: sounds arriving from the rear are attenuated above approximately 2~kHz, while frontal sounds retain high-frequency energy up to 8~kHz, approximating natural high-frequency shadowing by the head.

Each trajectory consists of multiple short tonal segments synthesized as harmonic violin-like waveforms with smooth attack and release envelopes to avoid transient artifacts. The base pitch of each segment is selected from a discrete musical scale, and higher harmonics are blended to maintain natural timbre while keeping spectral complexity minimal. The entire rendering process is deterministic and parameterized by sampling rate, ear distance, listener position, trajectory endpoints, motion speed, and noise level. This synthesis pipeline bridges auditory physics with model evaluation, enabling transparent analysis of spatial reasoning under controlled acoustics.

\subsection{Benchmark Composition and Variants}

The AudioMotionBench systematically covers all combinations of the eight canonical spatial positions, resulting in a total of $8\times7=56$ directed motion trajectories. Each trajectory corresponds to a physically simulated sound source traveling linearly from one position to another within the $6\,\text{m}\times6\,\text{m}$ spatial field, with the listener fixed at the center $(3.0,3.0)$. With an average clip duration of approximately $6\,\text{s}$ determined by the rendering pipeline, the full set of 56 trajectories yields a total of about $5.6$ minutes of spatial audio. This exhaustive pairing ensures balanced sampling across lateral, diagonal, and front–back motions, providing comprehensive coverage of the auditory motion space.

For each trajectory, binaural recordings are generated under four acoustic conditions: one clean (noise-free) and three levels of additive Gaussian noise corresponding to approximate signal-to-noise ratios of 35, 25, and 15~dB. Thus, every directional path yields four distinct versions of the same motion sequence, facilitating controlled robustness evaluation. In total, AudioMotionBench contains $56\times4=224$ synthesized audio clips.

\begin{figure}[t]
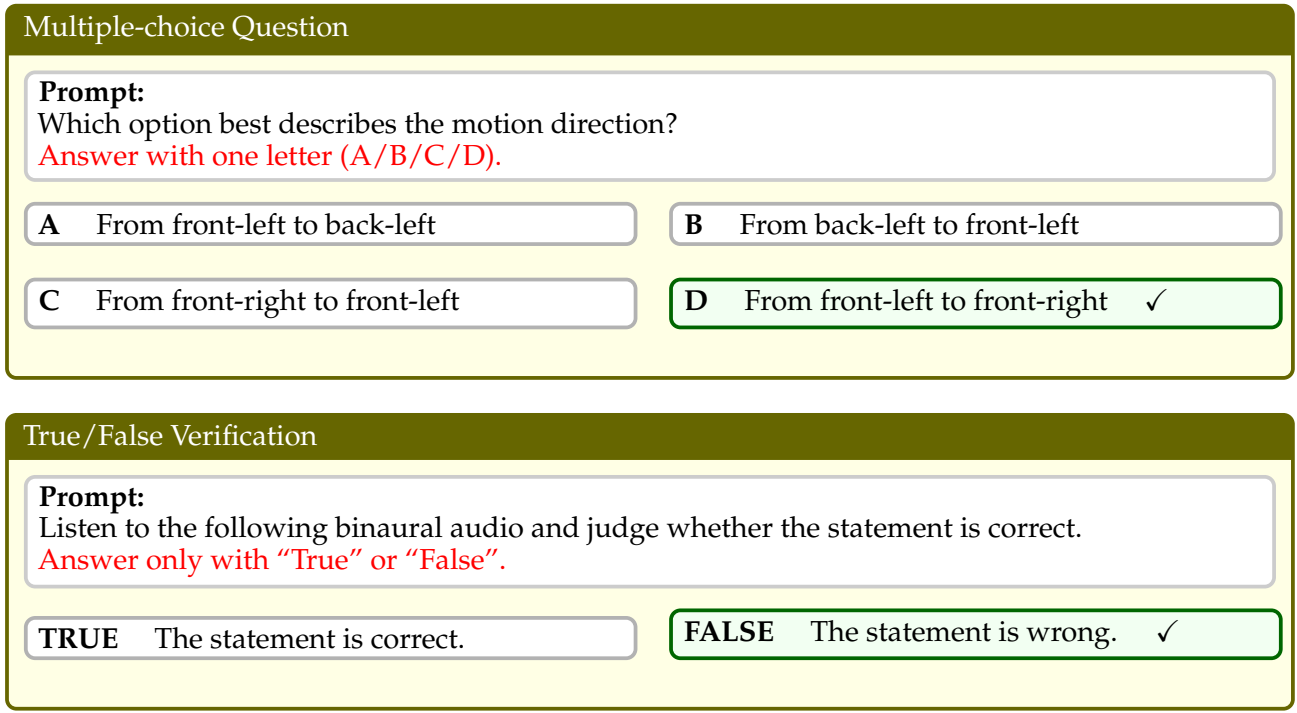

\centering

\renewcommand{\arraystretch}{1}
\def\OptionVSpace{\vspace{0.2mm}}

\begin{tcolorbox}[
    colback=yellow!10!white,
    colframe=yellow!40!black,
    title=Multiple-choice Question,
    left=1mm,right=1mm,top=1mm,bottom=1mm
]

\begin{tcolorbox}[
    colback=white,
    colframe=black!20,
    left=0.5mm,right=0.5mm,
    top=0mm,bottom=0mm
]
\textbf{Prompt:}\\
Which option best describes the motion direction?\\
\textcolor{red}{Answer with one letter (A/B/C/D).}
\end{tcolorbox}

\OptionVSpace

\begin{tabular}{@{}p{0.49\linewidth}p{0.49\linewidth}@{}}

\begin{tcolorbox}[colback=white,colframe=black!30,borderline={0.5pt}{0pt}{black!50,dashed},left=0.5mm,right=0.5mm,top=0mm,bottom=0mm]
\textbf{A} \quad From front-left to back-left
\end{tcolorbox}
&
\begin{tcolorbox}[colback=white,colframe=black!30,borderline={0.5pt}{0pt}{black!50,dashed},left=0.5mm,right=0.5mm,top=0mm,bottom=0mm]
\textbf{B} \quad From back-left to front-left
\end{tcolorbox}
\\

\begin{tcolorbox}[colback=white,colframe=black!30,borderline={0.5pt}{0pt}{black!50,dashed},left=0.5mm,right=0.5mm,top=0mm,bottom=0mm]
\textbf{C} \quad From front-right to front-left
\end{tcolorbox}
&
\begin{tcolorbox}[colback=green!5!white,colframe=green!40!black,borderline={0.5pt}{0pt}{green!40!black,dashed},left=0.5mm,right=0.5mm,top=0mm,bottom=0mm]
\textbf{D} \quad From front-left to front-right \quad \checkmark
\end{tcolorbox}
\\
\end{tabular}

\end{tcolorbox}

\OptionVSpace

\begin{tcolorbox}[
    colback=yellow!10!white,
    colframe=yellow!40!black,
    title=True/False Verification,
    left=1mm,right=1mm,top=1mm,bottom=1mm
]

\begin{tcolorbox}[colback=white,colframe=black!20,left=0.5mm,right=0.5mm,top=0mm,bottom=0mm]
\textbf{Prompt:}\\
Listen to the following binaural audio and judge whether the statement is correct.\\
\textcolor{red}{Answer only with ``True'' or ``False''.}
\end{tcolorbox}

\OptionVSpace

\begin{tabular}{@{}p{0.49\linewidth}p{0.49\linewidth}@{}}

\begin{tcolorbox}[colback=white,colframe=black!30,borderline={0.5pt}{0pt}{black!50,dashed},left=0.5mm,right=0.5mm,top=0mm,bottom=0mm]
\textbf{TRUE} \quad The statement is correct.
\end{tcolorbox}
&
\begin{tcolorbox}[colback=green!5!white,colframe=green!40!black,borderline={0.5pt}{0pt}{green!40!black,dashed},left=0.5mm,right=0.5mm,top=0mm,bottom=0mm]
\textbf{FALSE} \quad The statement is wrong. \quad \checkmark
\end{tcolorbox}
\\
\end{tabular}

\end{tcolorbox}

\caption{Example question formats in \textbf{AudioMotionBench}, including well-designed multiple-choice and true/false questions.}
\label{fig:ampbench_two_column}

\end{figure}

Each audio clip is paired with two types of language-based questions to probe different reasoning abilities. The first is a multiple-choice question, in which the model must identify the correct motion direction among four candidates (e.g., \texttt{A: Left$\rightarrow$Right}, \texttt{B: Right$\rightarrow$Left}, \texttt{C: Front$\rightarrow$Back}, \texttt{D: Back$\rightarrow$Front}). The second is a true–false verification question, which evaluates whether the model can judge the correctness of a given textual statement describing the motion. For every directional path, a pair of symmetric true–false statements is generated, doubling the total number of binary samples. As a result, the complete dataset contains 224 multiple-choice items and 448 true–false items, evenly distributed across the four noise levels.

To enable fine-grained analysis, AudioMotionBench further defines three controlled task variants probing complementary reasoning dimensions. The first variant, \emph{Spatial Consistency Task (SCT)}, keeps both frequency and velocity constant to isolate ITD/ILD-based spatial perception. The second, \emph{Spectral Disentanglement Task (SDT)}, varies the source pitch among seven discrete frequencies to test the ability to disentangle spatial cues from spectral interference. The third, \emph{Temporal Adaptation Task (TAT)}, maintains a fixed pitch while adjusting the motion velocity by a controllable scaling factor (e.g., doubled speed) to examine temporal adaptability and the continuity of motion reasoning. Together, these variants jointly assess the spatial, spectral, and temporal aspects of auditory reasoning under matched experimental protocols.

\begin{table}[h]
\centering
\label{tab:dataset_stats}
\begin{tabular}{m{1.2cm}@{\hskip 6pt}p{5.6cm}@{\hskip 12pt}p{3.4cm}}
\toprule
\textbf{Task} & \textbf{Description} & \textbf{Samples} \\
\midrule
SCT & Constant pitch and velocity. & 224 MCQ + 448 T/F \\
SDT & Discrete pitch variation. & 224 MCQ + 448 T/F \\
TAT & Fixed pitch, variable velocity. & 224 MCQ + 448 T/F \\
\bottomrule
\end{tabular}
\caption{Overview of AudioMotionBench task variants.}
\end{table}

By combining physical simulation, linguistic structure, and controlled variability, AudioMotionBench offers a comprehensive and interpretable testbed for auditory motion reasoning. It bridges low-level binaural cues with high-level question-answering evaluation—forming a foundation for the next generation of spatially aware audio–language models.

\section{Experiments and Results}
\label{sec:experiments}

\subsection{Experimental Setup}

\paragraph{Models and Evaluation Protocol.}
We evaluate four representative large audio–language models, including 
Qwen2-Audio-7B~\cite{chu2024qwen2}, Voxtral-Mini-3B-2507~\cite{liu2025voxtral}, Aero-1-Audio~\cite{mandel2023aero}, and Kimi-Audio-7B~\cite{ding2025kimi}, on the proposed AudioMotionBench. All models are tested in a zero-shot setting without any fine-tuning or adaptation. Each system receives binaural audio together with the motion-related question following the QA formats introduced in Section~\ref{sec:Benchmark}. Responses are normalized into canonical labels (\texttt{A–D} or \texttt{TRUE/FALSE}) using a rule-based parser, and any unrecognized output is counted as incorrect. Accuracy is averaged over three random seeds and across all task variants, following standard benchmark evaluation.

\paragraph{Decoding and Reporting Details.}
To reduce randomness from decoding, we repeat each evaluation with three different random seeds and average the results.  
Unless otherwise specified, we report mean accuracy and its standard deviation across seeds.  
All models are run with temperature $T=0$ for maximum determinism, and we keep decoding parameters (maximum length, stop tokens, etc.) consistent across tasks and models to ensure fair comparison.

\begin{table*}[t]
\centering
\renewcommand{\arraystretch}{1.25}
\setlength{\tabcolsep}{10pt}
\resizebox{\textwidth}{!}{
\begin{tabular}{c c c c c c c}
\hline
\textbf{Task} & \textbf{Model} & \textbf{clean} & \textbf{35db} & \textbf{25db} & \textbf{15db} & \textbf{AVG} \\
\hline
\multirow{4}{*}{\textbf{SCT}}
  & Qwen2-Audio-7B~\cite{chu2024qwen2} 
    & $40.48{\scriptstyle(\pm 0.12)}$
    & $42.26{\scriptstyle(\pm 0.07)}$
    & $\mathbf{40.48}{\scriptstyle(\pm 0.11)}$
    & $\mathbf{42.86}{\scriptstyle(\pm 0.06)}$
    & $\mathbf{41.52}{\scriptstyle(\pm 0.09)}$ \\
  & Voxtral-Mini-3B-2507~\cite{liu2025voxtral}        
    & $\mathbf{44.64}{\scriptstyle(\pm 0.10)}$
    & $37.50{\scriptstyle(\pm 0.13)}$
    & $39.88{\scriptstyle(\pm 0.14)}$
    & $39.29{\scriptstyle(\pm 0.08)}$
    & $40.33{\scriptstyle(\pm 0.12)}$ \\
  & Aero-1-Audio~\cite{mandel2023aero}    
    & $42.86{\scriptstyle(\pm 0.05)}$
    & $39.88{\scriptstyle(\pm 0.11)}$
    & $39.29{\scriptstyle(\pm 0.10)}$
    & $41.67{\scriptstyle(\pm 0.06)}$
    & $40.93{\scriptstyle(\pm 0.07)}$ \\
  & Kimi-Audio-7B~\cite{ding2025kimi}   
    & $43.45{\scriptstyle(\pm 0.15)}$
    & $\mathbf{43.45}{\scriptstyle(\pm 0.09)}$
    & $39.88{\scriptstyle(\pm 0.12)}$
    & $39.28{\scriptstyle(\pm 0.08)}$
    & $\mathbf{41.52}{\scriptstyle(\pm 0.11)}$ \\
\hline
\multirow{4}{*}{\textbf{SDT}}
  & Qwen2-Audio-7B~\cite{chu2024qwen2} 
    & $41.07{\scriptstyle(\pm 0.10)}$
    & $\mathbf{41.67}{\scriptstyle(\pm 0.06)}$
    & $\mathbf{44.64}{\scriptstyle(\pm 0.12)}$
    & $\mathbf{44.64}{\scriptstyle(\pm 0.07)}$
    & $\mathbf{43.01}{\scriptstyle(\pm 0.09)}$ \\
  & Voxtral-Mini-3B-2507~\cite{liu2025voxtral}    
    & $\mathbf{44.64}{\scriptstyle(\pm 0.14)}$
    & $39.88{\scriptstyle(\pm 0.10)}$
    & $41.67{\scriptstyle(\pm 0.08)}$
    & $38.69{\scriptstyle(\pm 0.13)}$
    & $41.22{\scriptstyle(\pm 0.12)}$ \\
  & Aero-1-Audio~\cite{mandel2023aero}   
    & $42.26{\scriptstyle(\pm 0.11)}$
    & $39.88{\scriptstyle(\pm 0.14)}$
    & $39.88{\scriptstyle(\pm 0.13)}$
    & $41.67{\scriptstyle(\pm 0.07)}$
    & $40.92{\scriptstyle(\pm 0.09)}$ \\
   & Kimi-Audio-7B~\cite{ding2025kimi}   
    & $42.86{\scriptstyle(\pm 0.08)}$
    & $39.88{\scriptstyle(\pm 0.07)}$
    & $38.69{\scriptstyle(\pm 0.15)}$
    & $39.26{\scriptstyle(\pm 0.10)}$
    & $40.17{\scriptstyle(\pm 0.11)}$ \\
\hline
\multirow{4}{*}{\textbf{TAT}}
  & Qwen2-Audio-7B~\cite{chu2024qwen2} 
    & $39.29{\scriptstyle(\pm 0.13)}$
    & $40.48{\scriptstyle(\pm 0.10)}$
    & $\mathbf{44.05}{\scriptstyle(\pm 0.11)}$
    & $42.64{\scriptstyle(\pm 0.12)}$
    & $41.62{\scriptstyle(\pm 0.09)}$ \\
  & Voxtral-Mini-3B-2507~\cite{liu2025voxtral}     
    & $\mathbf{44.05}{\scriptstyle(\pm 0.08)}$
    & $40.48{\scriptstyle(\pm 0.14)}$
    & $40.48{\scriptstyle(\pm 0.09)}$
    & $39.88{\scriptstyle(\pm 0.07)}$
    & $41.22{\scriptstyle(\pm 0.12)}$ \\
  & Aero-1-Audio~\cite{mandel2023aero}   
    & $42.86{\scriptstyle(\pm 0.06)}$
    & $39.88{\scriptstyle(\pm 0.07)}$
    & $39.88{\scriptstyle(\pm 0.11)}$
    & $41.67{\scriptstyle(\pm 0.08)}$
    & $41.07{\scriptstyle(\pm 0.10)}$ \\
   & Kimi-Audio-7B~\cite{ding2025kimi}   
    & $42.26{\scriptstyle(\pm 0.12)}$
    & $\mathbf{44.05}{\scriptstyle(\pm 0.09)}$
    & $38.10{\scriptstyle(\pm 0.13)}$
    & $\mathbf{42.86}{\scriptstyle(\pm 0.06)}$
    & $\mathbf{41.82}{\scriptstyle(\pm 0.11)}$ \\
\hline
\end{tabular}
}
\caption{
\textbf{Performance (\%) across three tasks and multiple noise levels.}
Each entry reports the mean accuracy and its standard deviation across repeated runs, enabling consistent comparison of model stability.
The best-performing system within each task and noise condition is highlighted in bold.
The table spans four noise levels (clean, 35\,dB, 25\,dB, and 15\,dB) as well as an averaged score (AVG), providing a comprehensive view of how model performance changes as the signal quality progressively degrades.
}
\label{tab:snr_results}
\end{table*}

\subsection{Benchmark Results and Analysis}

\paragraph{Overall Accuracy Trends.}
Table~\ref{tab:snr_results} summarizes accuracy across different motion tasks and signal-to-noise ratios (SNRs). Among the evaluated systems, Qwen2-Audio-7B and Kimi-Audio-7B achieve the strongest overall performance, with Voxtral-Mini-3B-2507 and Aero-1-Audio showing slightly lower but comparable results. However, none of the models surpass 50\% accuracy even under clean conditions, far below the level expected for reliable spatial perception.

\paragraph{Effect of Signal-to-Noise Ratio.}
Although we evaluated multiple models under varying signal-to-noise ratio (SNR) conditions, none of them exhibited noticeable performance changes as the SNR decreased. Across all models, accuracy varied by only a few percentage points when moving from clean audio to severely degraded ($15$dB) conditions.

Importantly, this observation should be interpreted within the specific context of auditory motion perception studied in AMPBench. The near-constant accuracy across SNR levels does not necessarily indicate general robustness to acoustic degradation; rather, it suggests that current LALMs fail to effectively leverage spatial cues whose reliability is expected to deteriorate under low-SNR conditions. As a result, reducing SNR does not further impair performance because the models already underutilize fine-grained spatial information embedded in the waveform.

This behavior contrasts sharply with human auditory perception, where motion discrimination performance degrades substantially as spatial cues such as ITD and ILD become unreliable in noisy environments.

\paragraph{Task-wise and Model-wise Differences.}
Across tasks, we observe that Task~2, which involves variable-speed motion, is slightly easier for most models than Task~1 and Task~3, indicating that some forms of variation may be easier to capture than front--back ambiguities. Model-wise, Qwen2-Audio-7B tends to dominate in Task~1 and Task~2 at higher SNRs, while Voxtral-Mini-3B-2507 occasionally achieves the best scores at clean condition, reflecting different robustness trade-offs between architectures. Nevertheless, the overall performance remains far from satisfactory in all regimes.

\paragraph{Task-format-specific Evaluation.}
Given this observation, we focus our analysis on the influence of task format by separately examining the accuracy on multiple-choice (MCQ) and true/false (T/F) questions. The corresponding accuracies are defined as follows:

\begin{equation}
\begin{aligned}
&\mathrm{Acc\text{-}MCQ} = \frac{1}{N}\sum_{i=1}^{N}\mathbb{1}\!\big[f_{\mathrm{MCQ}}(x_i)=y_i\big],
\end{aligned}
\end{equation}

\begin{equation}
\scalebox{0.9}{$
\mathrm{Acc\text{-}T/F} = \frac{1}{2N}\sum_{i=1}^{N}
\Big(\mathbb{1}[\hat{z}_i^{T}=\text{True}] +
     \mathbb{1}[\hat{z}_i^{F}=\text{False}]\Big).
$}
\end{equation}

where $f_{\mathrm{MCQ}}(x_i)$ denotes the predicted option for the $i$-th MCQ question, and $\hat{z}_i^{T}$ and $\hat{z}_i^{F}$ denote the model outputs when the same audio is paired with the True and False statements, respectively.
\begin{table*}[t]
\centering
\renewcommand{\arraystretch}{1.25}
\setlength{\tabcolsep}{7pt}

\resizebox{\textwidth}{!}{
\begin{tabular}{lcccccc}
\toprule
\multirow{2}{*}{Model} & 
\multicolumn{2}{c}{\textbf{SCT}} & 
\multicolumn{2}{c}{\textbf{SDT}} & 
\multicolumn{2}{c}{\textbf{TAT}} \\
\cmidrule(lr){2-3} \cmidrule(lr){4-5} \cmidrule(lr){6-7}
 & Acc-MCQ & Acc-T/F & Acc-MCQ & Acc-T/F & Acc-MCQ & Acc-T/F \\
\midrule

Qwen2-Audio-7B
& $24.77{\scriptstyle(\pm 0.12)}$ 
& $50.89{\scriptstyle(\pm 0.09)}$
& $25.34{\scriptstyle(\pm 0.11)}$
& $51.22{\scriptstyle(\pm 0.10)}$
& $25.88{\scriptstyle(\pm 0.13)}$
& $51.61{\scriptstyle(\pm 0.08)}$ \\

Voxtral-Mini-3B-2507
& $23.66{\scriptstyle(\pm 0.14)}$
& $\mathbf{52.68}{\scriptstyle(\pm 0.11)}$
& $24.13{\scriptstyle(\pm 0.12)}$
& $\mathbf{52.87}{\scriptstyle(\pm 0.09)}$
& $24.52{\scriptstyle(\pm 0.10)}$
& $\mathbf{53.11}{\scriptstyle(\pm 0.12)}$ \\

Aero-1-Audio
& $24.11{\scriptstyle(\pm 0.10)}$
& $50.45{\scriptstyle(\pm 0.07)}$
& $24.63{\scriptstyle(\pm 0.13)}$
& $50.88{\scriptstyle(\pm 0.12)}$
& $25.06{\scriptstyle(\pm 0.11)}$
& $51.34{\scriptstyle(\pm 0.10)}$ \\

Kimi-Audio-7B
& $\mathbf{25.05}{\scriptstyle(\pm 0.09)}$
& $51.26{\scriptstyle(\pm 0.08)}$
& $\mathbf{25.49}{\scriptstyle(\pm 0.10)}$
& $51.63{\scriptstyle(\pm 0.11)}$
& $\mathbf{25.97}{\scriptstyle(\pm 0.07)}$
& $52.02{\scriptstyle(\pm 0.09)}$ \\

\bottomrule
\end{tabular}
}
\caption{
\textbf{MCQ vs T/F accuracy on AudioMotionBench (clean).}
Each value reports the mean accuracy with its standard deviation across repeated runs. 
Bold numbers indicate the best-performing model within each task and question format.
}
\label{tab:mcq_tf_tasks}
\end{table*}

Table~\ref{tab:mcq_tf_tasks} summarizes the results under the clean condition of AMPBench. The accuracies of almost all models remain around 25\% for multiple-choice questions and 50\% for true/false questions across all three tasks. These values are almost exactly at the chance levels for four-way multiple choice (25\%) and balanced binary classification (50\%). This result indicates that current models possess little genuine discriminative ability in spatial auditory reasoning. The models appear to rely mainly on superficial spectral or linguistic cues rather than robust spatial representations such as interaural phase differences. Even the strongest models, which are trained on large-scale multimodal corpora, for instance Kimi-Audio-7B for MCQ and Voxtral-Mini-3B-2507 for T/F, yield at most a one- to two-point improvement over pure guessing, suggesting that using massive multimodal training data alone does not endow the models with spatial perception ability.

Overall, these findings reveal a persistent deficiency in spatial grounding across existing audio-language architectures. While they demonstrate strong semantic comprehension, they fail to model the geometric and directional structure of auditory scenes, leaving spatial perception an open challenge for future multimodal model design.

\subsection{Prompt Ablation}
\label{sec:prompt_ablation}

\begin{figure}[t]
\centering
\includegraphics[width=0.8\linewidth]{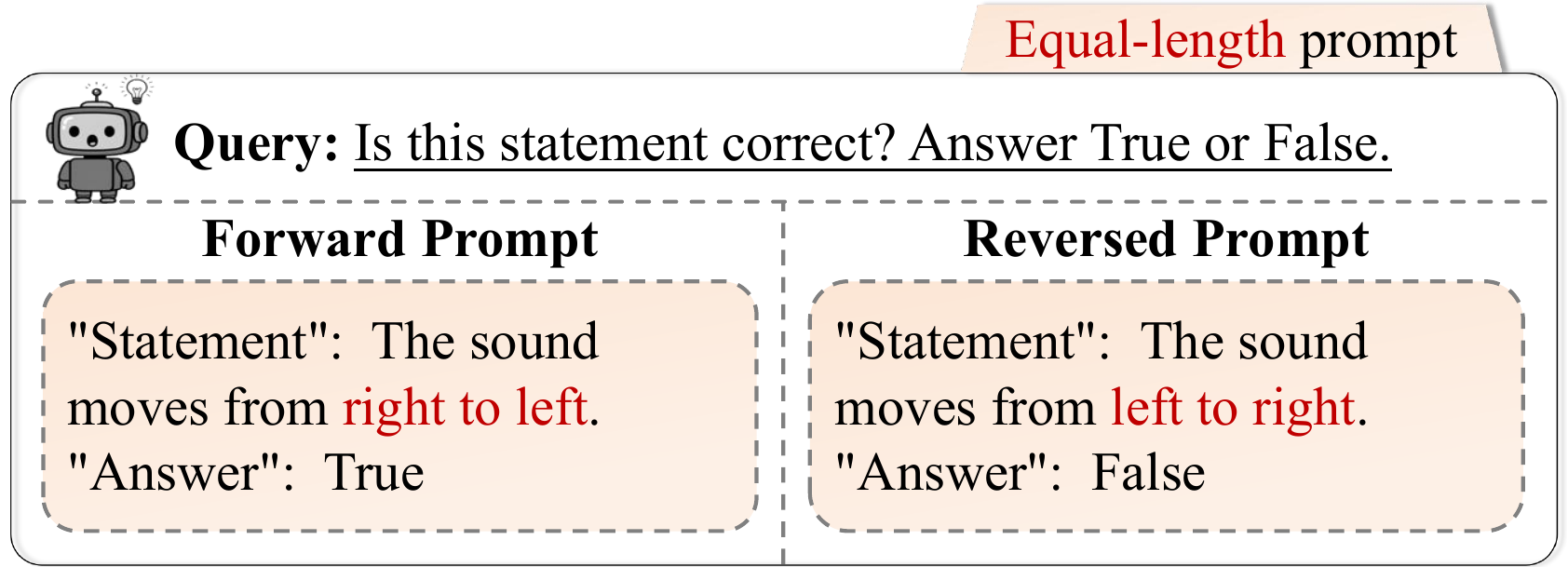}
\caption{\textbf{True/False prompt construction.} 
Each audio sample is paired with a True statement describing the ground-truth motion and a False statement created by mirroring the trajectory while matching lexical length and phrasing. }
\label{fig:tf_examples}
\end{figure}

To examine whether audio--language models genuinely verify auditory motion rather than following linguistic heuristics, we focus on the True/False (T/F) format of AudioMotionBench. Each audio sample is paired with two statements describing a motion trajectory. The \textit{True} statement corresponds to the ground-truth direction, whereas the \textit{False} statement is generated by mirroring the trajectory (e.g., left-to-right versus right-to-left) while matching lexical length, tense, and sentiment to eliminate textual shortcuts, as illustrated in Figure~\ref{fig:tf_examples}.

To quantify model reliability, we compute three complementary metrics:
the True-statement hit rate (\textbf{TPR}),
the False-statement rejection rate (\textbf{TNR}),
and the False→True misjudgment probability (\textbf{YesBias}):

\begin{equation}
\begin{aligned}
\mathrm{TPR}  & = \Pr(\hat{z}=\text{True}\mid z=\text{True}), \\
\mathrm{TNR}  & = \Pr(\hat{z}=\text{False}\mid z=\text{False}), \\
\mathrm{YesBias} & = \Pr(\hat{z}=\text{True}\mid z=\text{False}).
\end{aligned}
\end{equation}

Table~\ref{tab:tf_bias_mer} summarizes the results under the clean condition of AudioMotionBench.Across all evaluated models, a consistent behavioral pattern appears within the scope of our motion–verification setting. The hit rate on True statements (TPR) is notably low, with the strongest model, Aero-1-Audio, accepting fewer than $21\%$ of True statements. In contrast, the rejection rate for False statements (TNR) is considerably higher, with Kimi-Audio-7B exceeding $84\%$. This imbalance suggests that under our evaluation protocol, models tend to be conservative when deciding whether a statement matches the underlying motion.

\begin{table}[t]
\centering
\begin{tabular}{lccc}
\toprule
Model & TPR & TNR & YesBias \\
\midrule

Qwen2-Audio-7B     
& $18.30{\scriptstyle(\pm 0.11)}$
& $83.48{\scriptstyle(\pm 0.08)}$
& $16.52{\scriptstyle(\pm 0.09)}$ \\

Voxtral-Mini-3B-2507      
& $19.40{\scriptstyle(\pm 0.13)}$
& $80.34{\scriptstyle(\pm 0.10)}$
& $17.90{\scriptstyle(\pm 0.12)}$ \\

Aero-1-Audio  
& $\mathbf{20.32}{\scriptstyle(\pm 0.12)}$
& $78.93{\scriptstyle(\pm 0.09)}$
& $17.98{\scriptstyle(\pm 0.11)}$ \\

Kimi-Audio-7B  
& $18.85{\scriptstyle(\pm 0.14)}$
& $\mathbf{84.39}{\scriptstyle(\pm 0.10)}$
& $\mathbf{16.49}{\scriptstyle(\pm 0.08)}$ \\

\bottomrule
\end{tabular}

\caption{
\textbf{True/False verification on AudioMotionBench (clean).}
We report true positive rate (TPR), true negative rate (TNR), and YesBias (\%),
where lower YesBias indicates reduced tendency toward “Yes’’ responses.
Bold values mark the best performance for each metric.
}
\label{tab:tf_bias_mer}
\end{table}

YesBias offers a complementary perspective. Although Kimi-Audio-7B shows the lowest YesBias, it still accepts around $16.5\%$ of False statements, and other models remain close to $18\%$. These nontrivial acceptance rates indicate that the boundary between True and False is not sharply differentiated for motion-related inputs.

Taken together, the low TPR and comparatively high TNR indicate that, in this specific auditory–motion verification task, current models have difficulty confirming correct motion descriptions. Rather than consistently integrating binaural cues such as level or timing differences, the models appear to fall back on conservative decision strategies, making them more comfortable rejecting than affirming statements. 

Overall, these findings reveal a clear gap between the textual statements and the motion cues embedded in the audio. Although our evaluation focuses specifically on motion perception, the consistent difficulty in confirming True statements hints at a broader challenge: current audio–language models may have limited ability to integrate fine-grained spatial cues when performing grounded verification. While we do not claim that this limitation extends to all forms of auditory reasoning, the observed patterns suggest that motion-related grounding is not yet a reliably solved component of existing LALM systems.

\subsection{Discussion}
The findings above reveal several deep-rooted architectural bottlenecks that collectively underpin current LALMs’ inability to perceive auditory motion. First, most large-scale training corpora remain dominated by single-channel or pseudo-stereo recordings, which inherently lack key interaural cues such as interaural time difference (ITD) and interaural level difference (ILD). In the absence of these physically grounded signals, the models tend to encode incoming audio as flattened spectral templates rather than spatially structured wavefronts, making distinctions such as left--right and front--back fundamentally ambiguous and easily confounded.

Second, Transformer-based encoders, while highly effective for capturing temporal–spectral dependencies, provide no explicit mechanism for representing geometric transformations associated with moving sources. Standard self-attention treats each frame as an unordered token and therefore lacks the inductive bias needed to model smooth trajectories, spatial continuity, or coherent motion fields. As a result, the model’s internal representations remain largely insensitive to dynamic spatial structure, limiting its ability to track objects across time within the audio modality alone.

Third, the aggressive latent compression employed by many architectures, such as pooling entire clips into a single global embedding before passing them to the language decoder, tends to wash out spatial gradients altogether. Subtle but critical variations in binaural timing or intensity cues are often averaged away during this bottleneck, producing systematic left–right confusions and severely degrading depth perception.

Taken together, these shortcomings illustrate why modern LALMs may excel at interpreting ``what`` is heard but consistently fail at determining``where`` it originates or ``how`` it is moving. Bridging this gap will require moving beyond purely semantic supervision toward architectures and training paradigms that explicitly encode the physics of spatial hearing. Our experiments highlight several promising avenues for advancing spatial reasoning. One direction is to incorporate physically grounded acoustic features directly into the model, for instance explicit ITD/ILD estimators or differentiable binaural-rendering modules. Another is to leverage multi-microphone or ambisonic recordings during pre-training so that spatial cues become a primary learning signal rather than incidental background variation. Finally, employing joint objectives that couple motion prediction, source localization, and language-based reasoning may encourage models to develop richer, more geometrically coherent internal representations of auditory scenes. These emerging directions collectively point toward the next generation of spatially grounded audio–language models capable of perceiving motion as naturally as they describe it.

\section{Conclusion}
\label{sec:conclusion}

\paragraph{Summary of Findings.}
In our work, we introduce AudioMotionBench, a physically grounded benchmark designed to systematically evaluate auditory motion perception in large audio–language models (LALMs). Through controlled experiments and cross-model comparisons, we reveal a consistent deficiency in spatial reasoning: while current models can often recognize what is heard, they still struggle to perceive where the sound source is and how it moves. This gap appears across architectures of different scales and training regimes, suggesting a structural limitation rather than model-specific behavior.

\paragraph{Key Insights.}
Our analysis indicates that existing LALMs, which are predominantly trained on mono or pseudo-stereo audio, lack the representational capacity needed to encode interaural cues, temporal continuity, and geometric patterns that support reliable motion inference. Their internal representations therefore remain largely semantic and only weakly coupled to the physical structure of the auditory scene.

\paragraph{Broader Implications.}
Beyond serving as a diagnostic benchmark, AudioMotionBench provides a principled foundation for studying spatially grounded auditory intelligence. The ability to track moving sound sources is essential for real-world, embodied agents that must locate, follow, or respond to dynamic auditory events in their environment. AudioMotionBench thus highlights the need for richer training data, architectural elements that explicitly model spatial cues, and learning objectives that reward accurate reasoning about physical motion. We hope this work encourages further exploration toward audio–language systems capable of perceiving the world not only semantically but also spatially, enabling a more embodied form of auditory understanding.

\section{Limitations}
\label{sec:limitations}

While AudioMotionBench offers a controlled and fully reproducible testbed for studying auditory motion perception, it inevitably simplifies several aspects of real-world acoustic experience. The benchmark relies on analytically rendered spatial audio and a restricted set of motion patterns and sound types, which allows precise isolation of spatial cues but does not fully reflect the complexity of natural acoustic environments. In practice, factors such as reverberation, occlusion, interference, and diverse sound semantics may interact with motion perception in ways not captured by the current design.

As a result, performance on AudioMotionBench should be interpreted as a measure of a model’s ability to reason about idealized spatial motion cues, rather than a comprehensive assessment of real-world auditory intelligence. Extending the benchmark toward more realistic recordings, continuous motion dynamics, and richer sound categories remains an important direction for future work, building upon AudioMotionBench as an initial foundation.

\newpage

{
\small

\bibliographystyle{unsrtnat}
\bibliography{custom}
}
\clearpage

\end{document}